\documentclass[3p,times,twocolumn]{elsarticle}

\usepackage{ecrc}

\volume{00}

\firstpage{1}

\journalname{Nuclear and Particle Physics Proceedings}

\runauth{W.\ A.\ Horowitz}

\jid{nppp}

\jnltitlelogo{Nuclear and Particle Physics Proceedings}

\usepackage{amsmath}
\usepackage{amssymb}
\usepackage[T1]{fontenc}
\usepackage{subcaption}
\biboptions{numbers,square,sort&compress}
\usepackage{hyperref} 
\usepackage{color}

\definecolor{darkgreen}{rgb}{0,.7,0}
\definecolor{linkblue}{rgb}{0.,0.,0.9333}

\hypersetup{
    pdffitwindow=true,      
    pdfauthor={W. A. Horowitz},     
    pdfnewwindow=true,      
    bookmarksnumbered=true,
    colorlinks=true,       
    linkcolor=red,          
    citecolor=darkgreen,        
    filecolor=magenta,      
    urlcolor=cyan,           
    menucolor=blue,
    baseurl=http://www.arxiv.org/abs/,
}

\usepackage[figuresright]{rotating}

\newcommand{\reffig}[1]{Fig.~\ref{fig:#1}}
\newcommand{\refeq}[1]{Eq.~\ref{eq:#1}}

\newcommand{\fig}[1]{\reffig{#1}}
\newcommand{\eq}[1]{\refeq{#1}}

\newcommand{\infinity}{\infty}

\newcommand{\qhat}{\hat{q}}

\begin{document}

\begin{frontmatter}


\title{Time Dependent $\hat{q}$ from AdS/CFT}
\author{W.\ A.\ Horowitz}
\ead{wa.horowitz@uct.ac.za}
\ead[url]{http://webapp-phy.uct.ac.za/personal/horowitz/index.php}
\address{Department of Physics, University of Cape Town, Private Bag X3, Rondebosch 7701, South Africa}






\begin{abstract}
We present the first ever AdS/CFT calculation of $\qhat$ for a light quark jet as a function of position or, equivalently, time. Our result does not suffer from the gamma factor blow up of the usual time-independent AdS/CFT heavy quark setup and is qualitatively similar to, but differs by $\sim\mathcal{O}(1)$ factor from, the light flavor result of Liu, Rajagopal, and Wiedemann. Our findings can be immediately implemented into any $\qhat$-based energy loss model.

Our $\qhat$ derivation relies on our calculation of the average distance squared, $s^2(t)$, travelled by the endpoint of a string falling in an AdS$_3$-Schwarzschild spacetime. The early time behavior is ballistic, $s^2(t)\sim t^2$, but the late time behavior is the usual diffusive Brownian motion, $s^2(t)\sim t$. These late time dynamics are universal and depend only on the near-horizon physics, which allows us to generalize our results to arbitrary dimensions and thus make contact with the physics explored by RHIC and LHC.

Additionally, we find that AdS/CFT predicts angular ordering for radiation in medium, just as in vacuum, and in contradistinction to weak-coupling, with its anti-angular ordering prediction. Finally, our results also imply, sensibly, that AdS/CFT predicts a smooth interpolation between the angular correlations of open heavy flavor and light flavor observables.
\end{abstract}


\end{frontmatter}


\section{Introduction}
The goal of high energy heavy ion collisions is to understand the non-trivial emergent moderate body properties of non-Abelian quantum chromodynamics.  (Note that, since the number of ``particles'' interacting is on the order of 10K, one is possibly far from the thermodynamic limit; hence the use of ``moderate'' as opposed to ``many'' in describing the number of bodies involved.)  The most direct probes of the fundamental degrees of freedom and properties of the quark-gluon plasma produced in heavy ion collisions are the rare high momentum particles produced early in the collision that subsequently propagate out through the medium \cite{Majumder:2010qh}.  The original high-$p_T$ energy loss calculations assumed a weakly-coupled probe interacting with a weakly-coupled medium; i.e.\ that the dynamics of the medium and the interaction of the probe with the medium could be quantitatively described with the techniques of usual perturbative quantum field theory \cite{Majumder:2010qh}.  However, evidence from lattice QCD studies of the thermodynamic properties of QGP near the phase transition temperature and by comparing predictions from relativistic viscous hydrodynamics calculations to data from RHIC and LHC one finds that the soft sector of heavy ion collisions is probably best understood with the tools of strong coupling, most notably the AdS/CFT correspondence \cite{CasalderreySolana:2011us}.  Thinking more broadly, it is possible that the physics of high momentum probes is dominated by the soft, non-perturbative physics of the soft medium that is at a scale of $T\sim\Lambda_{QCD}$.  We are thus led to consider the possibility of a high momentum probe strongly coupled to a strongly coupled medium \cite{Gubser:2006bz,Herzog:2006se,Chesler:2008uy}.

\begin{figure}[!tbp]
        \centering
        \includegraphics[width=\columnwidth]{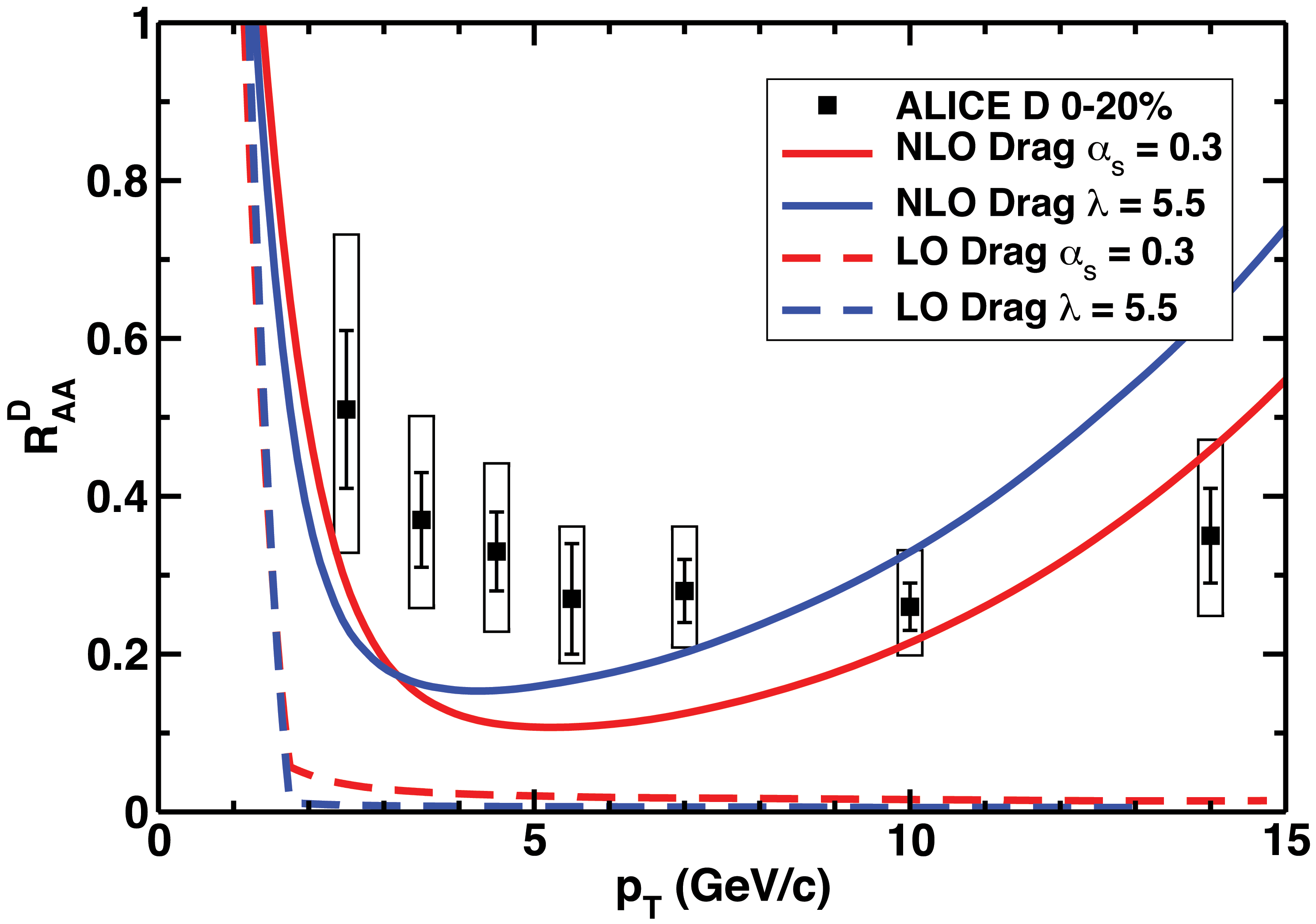}
    \caption{\label{fig:suppression}(Color online) $D$ meson $R_{AA}(p_T)$ measured by ALICE \protect\cite{ALICE:2012ab} compared to the strongly-coupled energy loss predictions with momentum fluctuations (NLO) and without fluctuations (LO) for two sets of reasonable QCD-like parameters in the $\mathcal{N}=4$ SYM theory \protect\cite{Horowitz:2015dta}.}
\end{figure}

\begin{figure}[!tbp]
\includegraphics[width=\linewidth]{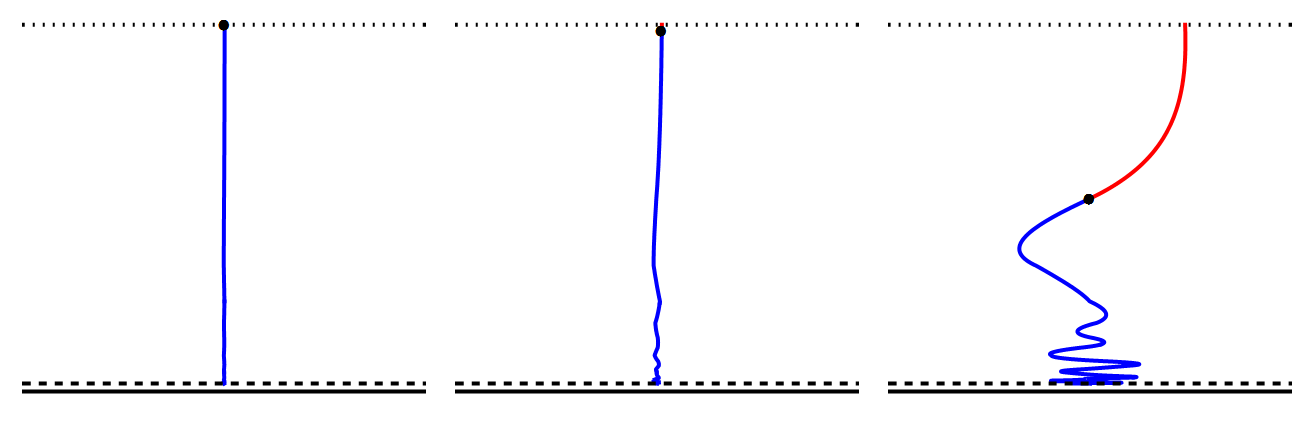}
\caption{(Color online) Temporal snapshots (with time increasing from left to right) of an equilibrating light quark string (blue curve) with the free endpoint (black dot) falling from its initial position towards the black hole horizon (dashed black line). The free endpoint of the light quark string can be interpreted as an observer travelling down the stretched string of de Boer et al.\ \protect\cite{deBoer:2008gu} (red curve) at the local speed of light.}
\label{fig:equilibratingstring}
\end{figure}

Previous calculations comparing the suppression of heavy and light flavor observables at leading order in AdS/CFT have shown mixed results.  In particular, naive applications of strong coupling calculations tend to oversuppress particles compared to data \cite{Morad:2014xla,Horowitz:2015dta}.  However, quantitative calculations of the suppression including the fluctuations in the mean momentum loss show much better qualitative agreement with data \cite{Akamatsu:2008ge,Horowitz:2015dta}; see \fig{suppression}.  This better agreement with data suggests that it is worthwhile to investigate further the fluctuations in energy loss in strongly coupled systems.  In particular, how does one go beyond the usual speed limit for heavy flavor physics? In a similar vein, since high-$p_T$ heavy quarks are like light quarks, what are the momentum fluctuations for light flavors?  These are very difficult questions, so we'll ask a simpler one: how do thermal fluctuations affect the motion of a light quark originally at rest in AdS/CFT?  See \fig{equilibratingstring} for a visualization of our setup.

\section{Brownian Motion of a Light Quark in AdS/CFT}
Even the above question is too hard to readily answer in general in the usual AdS$_5$ space.  We thus restrict ourselves to AdS$_3$ space in which we can invert the tortoise coordinates necessary for our calculation, which is based on the pioneering work of de Boer et al.\ \cite{deBoer:2008gu,Atmaja:2010uu} in which the Brownian motion of static heavy quarks was computed in AdS/CFT.

The outline of the solution for our problem is as follows: derive the classical solution for a straight, vertically stretched, originally static string whose one endpoint is fixed just above the black hole orizon and whose other endpoint away from the black hole horizon is allowed to freely fall in AdS$_3$-Schwarzschild; quantize the perpendicular directions of motion given the classical, leading order motion of the string; perform a semi-classical approximation by populating the quanta in the perpendicular direction according to Bose statistics; and compute the correlator related to the average distance squared position of the freely falling endpoint of the string.  One may see \cite{Moerman:2016wpv} for details.  The leading order classical solution is as one would expect: the free endpoint falls at the local speed of light.  Since we want to have a smooth interpolation between the heavy quark motion \cite{deBoer:2008gu} and the light flavor motion, we introduce a parameter $a$ that fixes as a fraction of the local speed of light the rate at which the free string endpoint is allowed to fall: $a=0$ corresponds to a heavy quark and $a=1$ to a light quark.  (We will also find having the ability to set $a(t)$ crucial in extracting $\qhat(t)$.) 

In general, the average distance squared position of the endpoint is a function of the plasma temperature, the initial length of the string (which corresponds to the initial virtuality of the quark), and $a$:
\begin{align}
	&s^2(t;a) 
	= \langle :\!\big(\hat{X}_{End}(t;a)-\hat{X}_{End}(0;a)\big)^2\!:\rangle \nonumber\\
	& = \frac{\beta^2}{4\pi^2\sqrt{\lambda}}\int_0^\infinity \frac{d\omega}{\omega}\frac{1}{e^{\beta\omega}-1}
	 |f_\omega(\sigma_f-at)-f_\omega(\sigma_f)e^{i\omega t}|^2,\nonumber
\end{align}
where $\beta=1/T$ is related to the temperature of the plasma, $\lambda$ is the 't Hooft constant, $\sigma_f$ is related to the initial length of the string, and $f_\omega$ are the Fourier modes supported by the string.

The above may be solved analytically in three scenarios: in the limit of small initial string length (small initial virtuality compared to the temperature); and for arbitrary virtualities at asymptotically early or late times, where the timescale dividing early from late times is set by the inverse temperature (regardless of the initial virtuality).

In the small virtuality limit one finds exactly
\begin{multline}
	s^2_\text{small}(t;a)=\frac{\beta^2}{4\pi^2\sqrt{\lambda}}\ln\left(\phantom{\frac{2 a \beta ^3 \sinh ^2\left(\frac{\pi  (a+1) t}{\beta }\right) \sinh ^2\left(\frac{\pi(a-1)t}{\beta }\right)  \text{csch}\left(\frac{2 \pi  a t}{\beta }\right)}{\pi ^3 \left(a^2-1\right)^2 t^3}}\right.\\
	\left.\frac{2 a \beta ^3 \sinh ^2\left(\frac{\pi  (a+1) t}{\beta }\right) \sinh ^2\left(\frac{\pi(a-1)t}{\beta }\right)  \text{csch}\left(\frac{2 \pi  a t}{\beta }\right)}{\pi ^3 \left(a^2-1\right)^2 t^3}\right).\label{eq:small-s}
\end{multline}
Since the above is somewhat difficult to analyze, it is worth expanding for small and late times (compared to the inverse temperature).  One finds that
\begin{align}
s_\text{small}^2(t;a)
&\xrightarrow{t\ll\beta} \frac{t^2}{6 \sqrt{\lambda}} + \mathcal{O}\big((t/\beta)^4\big) \label{eq:small-s-early} \\
&\xrightarrow{\beta\ll t} \frac{\beta t}{\pi  \sqrt{\lambda}}\left(1-\frac{a}{2}\right) + \mathcal{O}\big( \ln(\beta/t) \big).
\label{eq:small-s-late}
\end{align}
Thus the early time, small virtuality motion is ballistic while the late time, small virtuality motion is diffusive, with subdiffusive log corrections.

For arbitrary virtualities, one finds that
\begin{align}
s^2(t;a)
&\xrightarrow{t\ll\beta} c\frac{t^2}{\sqrt{\lambda}} + \mathcal{O}\big((t/\beta)^4\big) \label{eq:arb-early-s} \\
&\xrightarrow{\beta\ll t} s^2_\text{small}(t;a),
\label{eq:arb-late-s}
\end{align}
where $c$ is a numerical constant that depends on the initial virtuality of the string and $s^2_\text{small}(t;a)$ is precisely the small virtuality result, \eq{small-s}.  

The critically important point of \eq{arb-late-s} is that the late time dynamics of the fluctuating string are independent of the initial string length and solely determined by the near-horizon dynamics.  Since the tortoise coordinates are invertible near the black hole horizon in any number of dimensions, one may therefore extend the above results to AdS$_d$ for $d$ arbitrary.  One finds that
\begin{equation}
	s^2_{small}(t;a,d) = \frac{(d-1)^2}{4}s^2_{small}(t;a).
\end{equation}

Defining the diffusion coefficient in $d$ dimensions for a string whose endpoint is allowed to fall at a fraction $a$ of the local speed of light as $s^2(t;a,d)\equiv 2 D(a,d) t$, we find that
\begin{equation}
	D(a,d) = \frac{(d-1)^2\beta}{8\pi\sqrt{\lambda}}\left(1-\frac{a}{2}\right).
\end{equation}

Finally, we may relate $\qhat$ to the diffusion coefficient as follows.  $\qhat$ is defined as the transverse momentum squared imparted by the QGP to a high-$p_T$ particle per unit distance travelled.  In the usual AdS picture of a light quark pair evolving in a strongly coupled QGP (sQGP), a string forms a U-shape with the endpoints simultaneously falling and separating with time \cite{Chesler:2008uy,Morad:2014xla}.  If we concentrate on the transverse direction only, we may translate the falling, moving string endpoints to our late time endpoint falling (but stationary in all but the fifth dimension) at a fraction $a(t)$ of the local speed of light, which may be time dependent.  We then have that
\begin{align}
	\qhat 
	\equiv \frac{\langle p_T^2 \rangle}{\lambda_{mfp}} = \frac{2\kappa_T}{v} = \frac{4T^2}{vD} 
	= \frac{32\pi\sqrt{\lambda}T^3}{(d-1)^2(1-a/2)v}, \label{eq:qhat}
\end{align}
where the time dependence of $\qhat$ is absorbed into the time dependence of $a$ and $v$, the speed at which the light quark is propagating in the QGP.

\begin{figure*}[!tb]
\begin{subfigure}[t]{0.48\textwidth}
\centering
\includegraphics[width=\columnwidth]{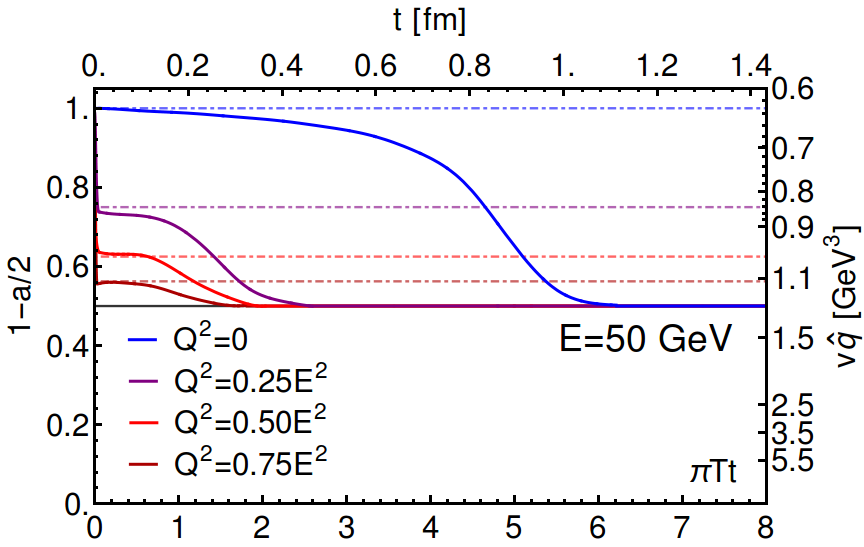}
\end{subfigure}
\hspace*{\fill}
\begin{subfigure}[t]{0.48\textwidth}
\centering
\includegraphics[width=\columnwidth]{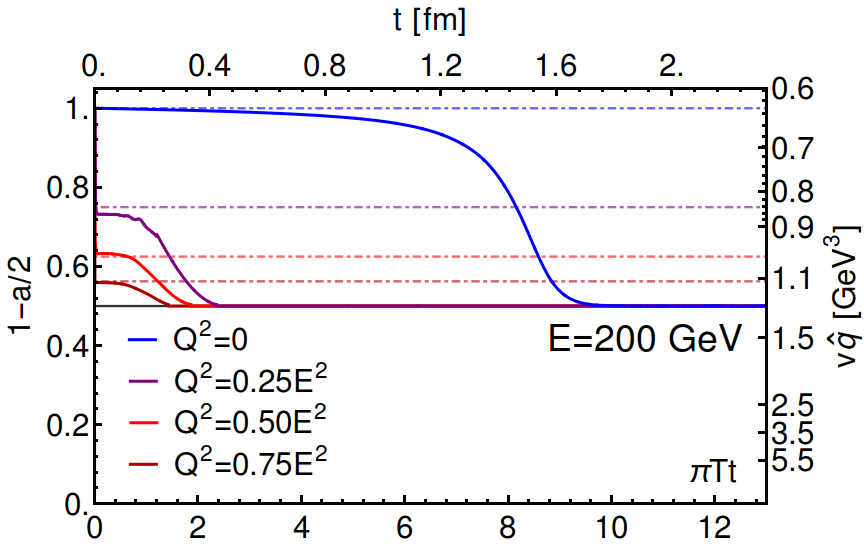}
\end{subfigure}
\\[.1in]
\hspace*{0.1in}%
\begin{subfigure}[t]{0.41\textwidth}
\centering
\includegraphics[width=\columnwidth]{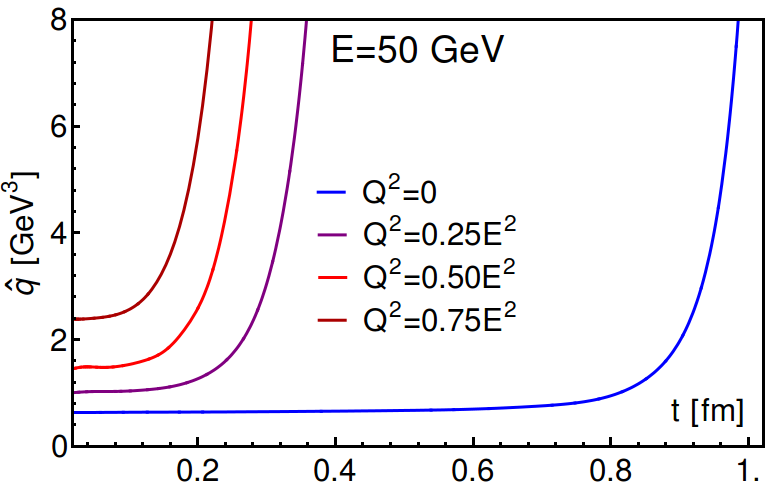}
\end{subfigure}
\hspace{.7in}%
\begin{subfigure}[t]{0.41\textwidth}
\centering
\includegraphics[width=\columnwidth]{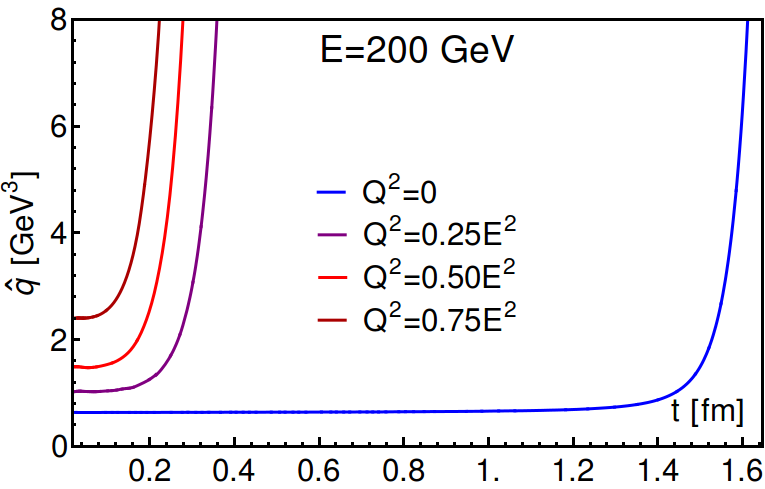}
\end{subfigure}
\hspace*{\fill}%
\caption{(Color online) Top row: $1-a/2\propto1/(v\times\hat{q})$ plotted as a function of $t$ for an $E =  50$ GeV quark (left) and $E = 200$ GeV quark (right) for various virtualities in a $T = 350$ MeV plasma with $\lambda = 5.5$. Bottom row: $\hat{q}$ plotted as a function of $t$ for an $E =  50$ GeV quark (left) and $E = 200$ GeV quark (right) for various virtualities in a $T = 350$ MeV plasma with $\lambda = 5.5$ \protect\cite{Moerman:2016wpv}.}
\label{fig:D}
\end{figure*}

\section{Conclusions and Outlook}
If we focus on $\qhat$ for heavy quarks, then $a(t) = 0$ and \eq{qhat} becomes
\begin{equation}
	\hat{q}_{MH}^{Heavy Quarks} = \frac{2\pi\sqrt{\lambda}T^3}{v}.
\end{equation}
Unlike the original fluctuations calculation \cite{Gubser:2006nz}, this result remains finite as $v\rightarrow1$; thus there is no speed limit associated with momentum fluctuations acting on the heavy quark.  What we conclude is that the rapid growth in $\gamma$ of the original fluctuations work is an artifact of the constant heavy quark velocity setup.  In this constant velocity setup, there is an infinite amount of energy stored in the string, which one can imagine will easily yield a ``tail wagging the dog'' situation in which the string fluctuations rapidly dominate the motion of the string endpoint.  Another way of thinking about the original Gubser momentum fluctuations is that perhaps they are due to the virtual particle pairs associated with the electric field on the D7 brane necessary to keep the heavy quark moving at a constant velocity; that the fluctuations are due to virtual field pairs would naturally explain their non-thermal nature.  

Turning our attention to light quarks, we find
\begin{equation}
	\hat{q}_{MH}^{Light Quarks} = \frac{2\pi\sqrt{\lambda}T^3}{(1-a/2)v}\simeq2\pi\sqrt{\lambda}T^3 \label{eq:lightqhat}
\end{equation}
for highly energetic quarks, whose endpoints only very slowly fall and whose velocities are nearly 1.  The above is numerically similar to the result from Liu, Rajagopal, and Wiedemann \cite{Liu:2006ug}, whose numerical coefficient is $\pi^{3/2}\Gamma(3/4)/\Gamma(5/4)\simeq7.5$ instead of our $2\pi\simeq6.2$.  

One may go further than the very large energy approximation used in \eq{lightqhat} and use numerical solutions for the U-shaped falling light quark--anti-quark pair string setup to determine $a(t)$ and $v(t)$.  We show the result of these calculations in \fig{D}.

One can see from the figure that $a$ is a monotonically increasing function of time (or, equivalently, distance travelled by the string endpoint). Thus the quark experiences its greatest transverse fluctuations at the initial time, and the fluctuations decrease monotonically with time.  This decrease in transverse fluctuations with time/distance is consistent with the pQCD result of angularly ordered gluon emissions from an off-shell parton in vacuum \cite{Mueller:1981ex,Ermolaev:1981cm} and in contradistinction to the anti-angular ordering observed in medium \cite{MehtarTani:2010ma}. 

Finally, since the transverse momentum fluctuations smoothly interpolate between light flavor and heavy flavors through $a$, one naturally expects a smooth interpolation between the angular correlations of $q\bar{q}$ pairs at high-$p_T$.

\section{Acknowledgments}
The author wishes to thank the South African National Research Foundation (NRF) and the SA-CERN Collaboration for generous support of this work.





\end{document}